\documentclass[12pt]{article}
\usepackage{rotating}
\usepackage{epsfig}
\usepackage{graphicx}
\usepackage{color}
\usepackage{cite}
\newcommand{\beq}{\begin{equation}}
\newcommand{\eeq}{\end{equation}}
\newcommand{\ga}{\lower.7ex\hbox{$\;\stackrel{\textstyle>}{\sim}\;$}}
\newcommand{\la}{\lower.7ex\hbox{$\;\stackrel{\textstyle<}{\sim}\;$}}

\begin{document}

\def\thefootnote{\fnsymbol{footnote}}

\begin{flushright}
{\tt KCL-PH-TH/2013-13}, {\tt LCTS/2013-22}, {\tt CERN-PH-TH/2013-156}  \\
{\tt ACT-6-13, MIFPA-13-22}\\
{\tt UMN-TH-3213/13,FTPI-MINN-13/23} \\
\end{flushright}

\begin{center}
{\bf {\Large Starobinsky-like Inflationary Models \\
\medskip
as Avatars of No-Scale Supergravity}}
\end{center}

\medskip

\begin{center}{\large
{\bf John~Ellis}$^{a}$,
{\bf Dimitri~V.~Nanopoulos}$^{b}$ and
{\bf Keith~A.~Olive}$^{c}$
}
\end{center}

\begin{center}
{\em $^a$Theoretical Particle Physics and Cosmology Group, Department of
  Physics, King's~College~London, London WC2R 2LS, United Kingdom;\\
Theory Division, CERN, CH-1211 Geneva 23,
  Switzerland}\\[0.2cm]
{\em $^b$George P. and Cynthia W. Mitchell Institute for Fundamental Physics and Astronomy,
Texas A\&M University, College Station, TX 77843, USA;\\
Astroparticle Physics Group, Houston Advanced Research Center (HARC), Mitchell Campus, Woodlands, TX 77381, USA;\\
Academy of Athens, Division of Natural Sciences,
28 Panepistimiou Avenue, Athens 10679, Greece}\\
{\em $^c$William I. Fine Theoretical Physics Institute, School of Physics and Astronomy,\\
University of Minnesota, Minneapolis, MN 55455, USA}\\[0.2cm]\end{center}

\bigskip

\centerline{\bf ABSTRACT}

\noindent  
Models of cosmological inflation resembling the Starobinsky $R + R^2$ model
emerge naturally among the effective potentials derived from no-scale 
SU(N,1)/SU(N) $\times$ U(1) supergravity when $N > 1$. We display several examples in
the SU(2,1)/SU(2) $\times$ U(1) case, in which the inflaton may be identified
with either a modulus field or a matter field. We discuss how the modulus field
may be stabilized in models in which a matter field plays the r\^ole of the inflaton. 
We also discuss models that generalize
the Starobinsky model but display different relations between the tilt in the spectrum
of scalar density perturbations, $n_s$, the tensor-to-scalar ratio, $r$, and the number
of e-folds, $N_*$. Finally, we discuss how such models can be probed by present and
future CMB experiments.
 
\begin{flushleft}
July 2013
\end{flushleft}
\medskip
\noindent

\newpage

\section{Introduction}

Although the first-year results from the Planck satellite~\cite{Planck} on the Cosmic Microwave Background (CMB)
are qualitatively consistent with generic expectations within the framework of cosmological inflation -
in particular, there are no signs of primordial non-Gaussianity in the CMB fluctuations or of
isocurvature perturbations, and the previous evidence for a tilt in the spectrum of scalar
perturbations, $n_s < 1$, has been confirmed - many simple inflationary models are challenged by
the Planck data - in particular, previous upper limits on the tensor-to-scalar ratio, $r$, have been
strengthened significantly. For example, single-field models with a monomial potential $\phi^n: n \ge 2$
are now disfavoured - at the $\sim 95$\% CL in the case of $\phi^2$ models, and at higher CLs for
models with $n > 2$. This has revived interest in non-monomial single-field potentials, such as
that found in the minimal Wess-Zumino model~\cite{CEM}~\footnote{Models with similar potentials
were proposed  long ago~\cite{ab} and more recently in~\cite{kltest}: see~\cite{oliverev} for a review.}.

The Planck constraints have also focused attention on the Starobinsky $R + R^2$ model,
which was proposed in 1980~\cite{Staro} and yields a spectrum of CMB perturbations that was analyzed
shortly afterwards by Mukhanov and Chibisov~\cite{MC}. 
The Starobinsky model yields a value of $n_s \sim 0.96$ that is in perfect agreement with the CMB data,
and a value of $r \sim 0.004$ that is comfortably consistent with the Planck upper limit~\cite{Planck}.

We take the point of view that cosmological inflation cries out for supersymmetry~\cite{ENOT}, in
the sense that it requires an energy scale that is hierarchically smaller than the Planck
scale, thanks to either a mass parameter being $\ll M_P$ and/or a scalar self-coupling
being $\ll {\cal O}(1)$. Since cosmology necessarily involves consideration of gravity,
it is natural to consider inflation in the context of local supersymmetry, i.e., supergravity~\cite{SUGRA}.
This preference is complicated, however, by the fact that a generic supergravity theory
has supersymmetry-breaking scalar masses of the same order as the gravitino mass, 
giving rise to the so-called $\eta$ problem~\cite{eta}, where the large vacuum energy density
during inflation leads to masses for all scalars of order the Hubble parameter \cite{glv}.
While inflationary models in simple supergravity can be constructed 
to avoid the $\eta$ problem \cite{nost,hrr}, these models rely on a seemingly
accidental cancellation in the inflaton mass \cite{lw}.

For this reason, we have long advocated no-scale supergravity~\cite{no-scale,EKN1,EKNGUT,LN} as the natural framework for
constructing models of inflation~\cite{EENOS,othernoscale,BG}. We have recently revived this proposal in light of the
Planck data, constructing an SU(2,1)/SU(2) $\times$ U(1) no-scale version of the minimal 
Wess-Zumino model~\cite{ENO6}~\footnote{For an alternative supergravity incarnation of the
Wess-Zumino inflationary model, see~\cite{Yanagida}.}.
We have shown that this NSWZ model is consistent with the Planck data for a
range of parameters that includes a special case in which it reproduces {\it exactly}
the effective potential and hence the successful predictions of the Starobinsky $R + R^2$ model~\cite{ENO6}.
We learnt subsequently that the $R + R^2$ model had previously been recovered from another
version of no-scale SU(2,1)/SU(2) $\times$ U(1) supergravity~\cite{Cecotti}, in a paper that makes deep observations on
connections between no-scale supergravity and higher-order gravity theories, including attractive
properties beyond the quadratic level, though without making the connection with
cosmology~\footnote{Subsequent to our paper, other Starobinsky
avatars of no-scale supergravity has been proposed and their implications for inflation investigated~\cite{KLno-scale,WB,FKR}. For other approaches to the embedding of higher-order gravity in the 
context of supergravity see~\cite{ketov}.}.
We note also that Higgs-inflation models~\cite{HI} and certain models with
conformally coupled fields~\cite{others} yield predictions similar to the $R + R^2$ model.

In this paper we discuss more generally avatars of no-scale supergravity that reproduce
the effective potential of the Starobinsky $R + R^2$ model, as well as related models that
yield similar predictions for the CMB.

As we show in Section~2 of this paper, the conformally-equivalent formulation of the Starobinsky model
in terms of a scalar field $\varphi$ has a kinetic term that is identical with that of the scalar
sector in the minimal no-scale SU(1,1)/U(1) supergravity model, reflecting a basic scaling
property of the underlying K\"ahler metric. However, we find no choice of the
superpotential for the SU(1,1)/U(1) model that can reproduce the effective scalar potential of the Starobinsky model.
On the other hand, we show in Section~3 that there are many possible choices of the
superpotential for the next-to-minimal SU(2,1)/SU(2) $\times$ U(1) no-scale supergravity
model that reproduce the Starobinsky potential, generalizing the examples previously
displayed in~\cite{ENO6}, \cite{Cecotti} and~\cite{KLno-scale,WB,FKR}. The corresponding K\"ahler metric inherits
the scaling property of the SU(1,1)/U(1) model that mimics the Starobinsky model, and is
parametrized by two complex fields, one of which could correspond to a modulus of a
string compactification and the other to a generic matter field. Some of the choices of
superpotential yield models in which the Starobinsky scalar field is identified with the
modulus field, and some with the matter field. In the latter case, the question arises
how the modulus field is stabilized. In Section~4 we give examples showing that
stabilization can be achieved without affecting the correspondence with the Starobinsky model.
Section~5 contains a discussion of models that resemble this model, yielding similar
predictions for the CMB observables. We discuss the extent to which these models
are constrained by the Planck and other data, and how future data could discriminate
further between the Starobinsky and other models. Finally, Section~6 summarizes our conclusions.

\section{The Starobinsky Model and No-Scale Supergravity}

Starobinsky considered in 1980~\cite{Staro} a generalization of the Einstein-Hilbert action to contain an $R^2$ contribution,
where $R$ is the scalar curvature:
\begin{equation}
S=\frac{1}{2} \int d^4x \sqrt{-g} (R+\alpha R^2) \, ,
\label{Staro}
\end{equation}
where $M \ll M_P$ is some mass scale. As was shown by Stelle in 1978~\cite{Stelle} and by Whitt in 1984~\cite{whitt}, 
the theory (\ref{Staro}) is conformally equivalent to a theory combining canonical gravity with a scalar field $\varphi$,
described by
\begin{equation}
S=\frac{1}{2} \int d^4x \sqrt{-g} \left[(1 + 2\alpha \varphi) R - \alpha \varphi^2 \right] \, ,
\label{Whitt}
\end{equation}
as can be seen trivially using the Lagrange equation for $\varphi$ in (\ref{Whitt}).
Making the Weyl rescaling $\tilde{g}_{\mu\nu} = (1 + 2 \alpha \varphi) g_{\mu\nu}$,
equation (\ref{Whitt}) takes the form
\begin{equation}
S=\frac{1}{2} \int d^4x \sqrt{-g} \left[ R + \frac{6 \alpha^2 \partial^\mu \varphi \partial_\mu \varphi}
{(1 + 2 \alpha \varphi)^2} - \frac{\alpha \varphi^2}{(1 + 2 \alpha \varphi)^2} \right] \, .
\label{Cecotti4}
\end{equation}
Making now the field redefinition $\varphi^\prime = \sqrt{\frac{3}{2}} \ln \left( 1+ \frac{\varphi}{3 M^2} \right)$ with $\alpha = 1/6M^2$,
one obtains a scalar-field action with a canonical kinetic term:
\begin{equation}
S=\frac{1}{2} \int d^4x \sqrt{-\tilde{g}} \left[\tilde{R} + (\partial_\mu \varphi^\prime)^2 - \frac{3}{2} M^2 (1- e^{-\sqrt{2/3}\varphi^\prime})^2 \right] \, ,
\end{equation}
in which the scalar potential takes the form
\begin{equation} 
V =  \frac{3}{4} M^2 (1- e^{-\sqrt{2/3}\varphi^\prime})^2 \, .
\label{r2pot}
\end{equation}
The spectrum of cosmological density perturbations found by using (\ref{Staro})
for inflation were calculated by Mukhanov and Chibisov in 1981~\cite{MC} and by
Starobinsky in 1983~\cite{Staro2}. The current data on cosmic microwave
background (CMB) fluctuations, in particular those from the Planck satellite~\cite{Planck},
are in excellent agreement with the predictions of this $R + R^2$ model.

As a preliminary to our comparison with no-scale supergravity, we first recall some general
features of the effective low-energy theory derived from a generic supergravity theory.
Neglecting gauge interactions, which are inessential for our purposes, any such theory
is characterized by a K\"ahler potential $K(\phi_i,  \phi^*_j)$, which is a hermitian function
of the chiral fields $\phi_i$ and their conjugates $\phi^*_j$, and a superpotential $W(\phi_i)$,
which is a holomorphic function of the $\phi_i$, via the combination
$G \equiv K + \ln W + \ln W^*$.  The effective field theory contains a generalized kinetic energy term
\begin{equation}
{\cal L}_{KE} \; = \; K^{ij^*} \partial_\mu \phi_i \partial \phi^*_j \, ,
\label{LK}
\end{equation}
where the K\"ahler metric $K^{ij^*} \equiv \partial^2 K / \partial \phi_i \partial \phi^*_{j}$, and the
effective scalar potential is
\begin{equation}
V \; = \; e^G \left[ \frac{\partial G}{\partial  \phi_i} K_{ij^*}  \frac{\partial G}{\partial  \phi^*_j} - 3 \right] \, ,
\label{effpot}
\end{equation}
where $K_{ij^*}$ is the inverse of the K\"ahler metric.

In parallel to the developments in the Starobinsky model described above,
the early 1980s were also the period when no-scale supergravity was discovered~\cite{no-scale},
developed and applied to particle phenomenology~\cite{EKN1,EKNGUT}, and subsequently derived from simple
compactifications of string theory~\cite{Witten} and proposed as a framework for constructing models of inflation~\cite{EENOS}.
The minimal no-scale SU(1, 1)/U(1) model may be written in terms of a single complex
scalar field $T$ with the K\"ahler function
\begin{equation}
K \; = \; -3 \ln ( T + T^*) \, .
\label{K3}
\end{equation}
In this case, the kinetic term becomes
\begin{eqnarray}
{\cal L}_{KE} \; = \;  \frac{3}{(T + T^*)^2} \partial_\mu T^*  \partial^\mu T \, ,
\label{no-scaleLKE}
\end{eqnarray}
and the effective potential becomes
\begin{equation}
V \; = \; \frac{{\hat V}}{(T + T^*)^2} \, : \, {\hat V} \; = \; \frac{1}{3}(T + T^*) |W_T|^2 - (W W_T^* + W^* W_T)\, .
\label{effV1}
\end{equation}
Generalizations including more chiral fields are described in the next Section.

For convenience, we recall here the action of the SU(1,1) group of isometric
transformations on the field $T$~\cite{EKN1}:
\begin{equation}
T \; \to \; \frac{ \alpha T + i \beta}{i \gamma T + \delta}: \; \alpha, \beta, \gamma, \delta ~{\rm real}, \; \alpha \delta + \beta \gamma \; = \; 1 \, .
\label{SU11}
\end{equation}
We exhibit explicitly the following SU(1,1) transformations:

$\bullet$ Imaginary translations:
\begin{equation}
T \; \to \; T + i \beta \, ,
\label{beta}
\end{equation}
under which the K\"ahler function $K = -3 \ln ( T + T^*)$ is invariant, but {\it not} the superpotential, in general.

$\bullet$ Dilatations:
\begin{equation}
T \; \to \; \alpha^2 T \, ,
\label{alpha}
\end{equation}
under which {\it neither} the K\"ahler function $K$ {\it nor} the superpotential is invariant,
whereas the no-scale kinetic term (\ref{no-scaleLKE}) {\it is} invariant under the transformation (\ref{alpha}).

$\bullet$ Conformal transformations:
\begin{equation}
\Delta T \; = \; - i \tau \left( \frac{T^2 - 1}{1 + i \tau T } \right) \, ,
\label{tau}
\end{equation}
under which again {\it neither} the K\"ahler function $K$ {\it nor} the superpotential is invariant.

$\bullet$ Inversions:
\begin{equation}
T \; \to \; \left( \frac{\beta}{\gamma} \right) \frac{1}{T} \, ,
\label{invert}
\end{equation}
under which the K\"ahler potential remains invariant, but the superpotential $W \to T \times W$.

The complex chiral field $T \equiv (t + i u)/\sqrt{2}$ parametrizes the non-compact two-dimensional coset space SU(1,1)/U(1),
the phase transformation $T \to T e^{i \theta}$ being equivalent to (\ref{beta}): $u \to u + i \theta$ for small $\beta$ and $\theta$.

We now note the obvious correspondence between the kinetic terms for the conformal scalar field in the 
Starobinsky model (\ref{Cecotti4}) and the no-scale field in (\ref{no-scaleLKE}),
once we make the identification $ (1 + \alpha \varphi) \leftrightarrow t$. This identity
reflects the partial invariance of both theories under the non-compact U(1) scale transformations: $t \; \to \; \alpha^2 t$ (\ref{alpha}),
and the analogous transformation for the scalar kinetic term in the Starobinsky model (\ref{Cecotti4}).

In general, neither of the effective potentials in the Starobinsky model and the no-scale SU(1,1)/U(1) model
is invariant under this rescaling of the corresponding scalar field.
However, in the case of the Starobinsky model this invariance under non-compact U(1) scaling is restored in the limit of large $\varphi$,
and the invariance of the effective potential at large $\varphi$ with a non-zero value yields inflation.
The scaling is broken explicitly by a term that is ${\cal O}(1/\varphi)$, which determines the slow-roll parameters.

The natural question then arises how such an inflationary potential may also arise for the $t$ field component in no-scale SU(1,1)/U(1)
supergravity. Looking at the form (\ref{effV1}) of the effective potential in the case, we see that
{\it iff} the superpotential $W \sim T^{3/2}$ at large $T$ the desired
scaling invariance of $V$ would be obtained. In this case the reduced
potential ${\hat V} \sim t^2$ at large $t$, a dependence cancelled by the denominator in $V = {\hat V}/2t^2$.
However, even setting aside the question whether such an asymptotic behaviour of $W$ can be made
compatible with holomorphy requirements, it is easy to check that the coefficient of the leading term at large $t$ would be {\it negative}:
\begin{equation}
W \; \sim \; A T^{3/2} + \dots \; \to \; {\hat V} \; \sim - \frac{3}{4} A^2 t^2 + \dots \, ,
\label{negative}
\end{equation}
so that Starobinsky inflation is impossible in this simplest no-scale SU(1,1)/U(1)
supergravity model. Accordingly, in the next Section we explore the possibilities in the simplest
non-minimal no-scale supergravity model.

\section{Obtaining the Starobinsky Model from SU(2,1)/SU(2) $\times$ U(1) No-Scale Supergravity}

We consider a no-scale supergravity model with two complex fields $(T, \phi)$ that
parametrize the non-compact SU(2,1)/SU(2) $\times$ U(1) coset space. In this case, the
K\"ahler potential may be written in the form
\begin{equation}
K \; = \; - 3 \ln \left(T + T^* - \frac{\phi \phi^*}{3} \right) \, ,
\label{K21}
\end{equation}
which has the obvious extension to SU(N,1)/SU(N) $\times$ U(1) models with $N - 1$ fields $\phi_i$~\cite{EKNGUT}.
Within this parameterization and the context of string compactification, the field $T$ has the natural interpretation
as a volume modulus, and $\phi$ as a generic matter field.
The K\"ahler potential (\ref{K21}) yields the following kinetic terms
for the scalar fields $T$ and $\phi$:
\begin{eqnarray}
{\cal L}_{KE} \; =  \; \left( \partial_\mu \phi^*, \partial_\mu T^* \right) \left(\frac{3}{(T + T^* - |\phi|^2/3)^2} \right)
 \left( \begin{array}{cc} (T + T^*)/3 & - \phi \\ - \phi^* & 1 \end{array} \right)
\left( \begin{array} {c} \partial^\mu \phi \\ \partial^\mu T \end{array} \right) \, .
\label{no-scaleL}
\end{eqnarray}
For a general superpotential $W(T,\phi)$, the
effective potential becomes
\begin{equation}
V \; = \; \frac{{\hat V}}{(T + T^* - |\phi|^2/3)^2} 
\label{VVhatT}
\eeq
with
\beq
{\hat V} \; \equiv \; \left| \frac{\partial W}{\partial \phi} \right|^2  +\frac{1}{3} (T+T^*) |W_T|^2 +
\frac{1}{3} \left(W_T (\phi^* W_\phi^* - 3 W^*) + {\rm h.c.}  \right) \, ,
\label{effV}
\end{equation}
where $W_\phi = \partial W/\partial \phi$ and $W_T = \partial W/\partial T$.
In early no-scale models of inflation~\cite{EENOS,BG} it was assumed that $K$ was
fixed, i.e., that the combination $(T + T^* - |\phi|^2/3)$ was fixed, and $W$ was a function of
$\phi$ only, so that the potential was simply $\hat{V} = |W_\phi|^2$
up to a trivial re-scaling. More recently, we assumed~\cite{ENO6} that the $T$ field was fixed, with a
vacuum expectation value (vev) $2 \langle Re T \rangle = c$ and $\langle Im T \rangle = 0$
that was determined by some unspecified non-perturbative high-scale dynamics~\footnote{For previous
proposals how this might occur, see the KKLT~\cite{KKLT} and KL models~\cite{KL,fixT}.}.
It was shown that in such a case the Starobinsky inflationary potential for $\phi$ would be obtained
with the following Wess-Zumino choice of superpotential:
\begin{equation}
W \; = \; \frac{\hat \mu}{2} \Phi^2 - \frac{\lambda}{3} \Phi^3 \, .
\label{WZW}
\end{equation}
and $\lambda = \mu/3$ where $\mu = {\hat \mu} / \sqrt{c/3}$.

Here we adopt an agnostic approach, starting from a more symmetric representation of
the SU(2,1)/SU(2) $\times$ U(1) coset space~\cite{EKNGUT}:
\begin{equation}
K \; = \; - 3 \ln \left(1 - \frac{|y_1|^2 + |y_2|^2}{3} \right) \, ,
\label{K21symm}
\end{equation}
where the complex fields $y_{1,2}$ are related to the fields $T, \phi$ appearing in (\ref{K21}) by
\begin{equation}
y_1 \; = \; \left( \frac{2 \phi}{1 + 2 T} \right) \; ; \; y_2 \; = \; \sqrt{3} \left( \frac{1 - 2 T}{1 + 2 T} \right) \, ,
\label{Tphiwrite}
\end{equation}
with the inverse relations
\begin{equation}
T \; = \; \frac{1}{2} \left( \frac{1 - {y_2}/{\sqrt{3}}}{1 + {y_2}/{\sqrt{3}}} \right)\, ; \;
\phi \; = \; \left(\frac{y_1}{1 + {y_2}/{\sqrt{3}}} \right) \, .
\label{Tphirewrite}
\end{equation}
When the coordinates are transformed as in (\ref{Tphiwrite}, \ref{Tphirewrite}), the
effective superpotential is modified:
\begin{equation}
W(T, \phi) \; \to \; {\widetilde W}(y_1, y_2) \; = \; \left( 1 + {y_2}/{\sqrt{3}} \right)^3 W  \, .
\label{Wtilde}
\end{equation}
For convenience, in the following we drop the tilde over the superpotential, and consider
various superpotentials $W(y_1, y_2)$ that yield an effective Starobinsky inflationary potential.

For convenience, we first provide some general formulae that provide a framework for
the specific examples discussed below. In a generic model specified by
\begin{equation}
G \; = \; - 3 \ln \left(1 - \frac{|y_1|^2 + |y_2|^2}{3}\right) + \ln |W|^2 \, ,
\label{general}
\end{equation}
one has an effective potential
\begin{equation}
V \; = \; \frac{{\hat V}}{(1 - (|y_1|^2 + |y_2|^2)/3)^2} \, ,
\label{Vhatphii}
\end{equation}
where
\begin{eqnarray}
{\hat V} & = & (1 - |y_1|^2/3) |W_1|^2 + (1 - |y_2|^2/3) |W_2|^2
- 3 |W|^2 \nonumber \\
& + & \left( (y_1 W_1 + y_2 W_2) W^* - \frac{y_1 y_2^*}{3} W_1 W_2^* + {\rm h.~c.} ) \right) \, ,
\label{fullVhat}
\end{eqnarray}
where $W_{1,2} = \partial W/\partial y_{1,2}$.
If one now sets, for example, $\langle y_2 \rangle = 0$, one finds
\begin{eqnarray}
V & = & \frac{{\hat V}}{(1 - |y_1|^2/3)^2} : \nonumber \\
{\hat V} & = & (1 - |y_1|^2/3) |W_1|^2 + |W_2|^2 - 3 |W|^2 +
(y_1 W_1 W^* + {\rm h.~c.}) \, ,
\label{useful}
\end{eqnarray}
and the dynamical field $y_1$ can be converted into a canonically-normalized inflaton field $x$ by the transformation
\begin{equation}
y_1 \; = \; \pm \sqrt{3} \tanh (\chi/\sqrt{3} )\; = \; \pm  \sqrt{3} \tanh (x/\sqrt{6} ) \, ,
\label{arctanh}
\end{equation}
where $\chi = (x + iy)/\sqrt{2}$ and the latter equality holds for $y=0$.

Before we describe some more details of the construction of SU(2,1)/SU(2) $\times$ U(1) no-scale inflationary models,
we note that there are two general forms for the potential that we are searching for.
First, recall the form of the kinetic term and potential in Eqs. (\ref{no-scaleLKE}) and (\ref{effV1}) for the modulus $T$.
The potential (\ref{r2pot}) is found when 
\beq
\hat{V} = 3M^2 |T-1/2|^2 \, ,
\label{gen1}
\eeq
which yields a potential $V$ that is independent of $T$ in the limit of large $T$, 
and hence invariant asymptotically under the dilatation transformation (\ref{alpha}), 
as can be seen using (\ref{VVhatT}).
We can obtain a canonically-normalized kinetic term by making the field redefinition
\beq
T \; = \; \frac{1}{2}e^{2\chi/\sqrt{3}} \, ,
\eeq
for which the Lagrangian becomes
\beq
{\cal L} \; = \; {\rm sech}^2((\chi-\chi^*)/\sqrt{3}) |\partial_\mu \chi|^2 - 12 M^2 \frac{e^{(\chi + \chi^*)/\sqrt{3}}}{
 (e^{2\chi/\sqrt{3}}+e^{2\chi^*/\sqrt{3}})^2}|\sinh(\chi/\sqrt{3})|^2 \, .
 \label{leff}
\eeq
Writing $\chi$ in terms of its real and imaginary parts: $\chi = (x + iy)/\sqrt{2}$, this becomes
\begin{eqnarray}
\label{leff2}
&& {\cal L}  \; = \; \frac{1}{2}\sec^2(\sqrt{2/3} y) \left( (\partial_\mu x)^2 + (\partial_\mu y)^2 \right) - \\
& & \,\, 3 M^2 \frac{e^{-\sqrt{2/3}x}}{2} \sec^2(\sqrt{2/3}y)\left(\cosh{\sqrt{2/3}x})-\cos{\sqrt{2/3}y}\right) \, , \nonumber
\end{eqnarray}
which reduces to (\ref{r2pot}) when $\langle y \rangle = 0$ for the canonical field $x$.

Note that the same potential can also be obtained if
\beq
\hat{V} = 12 M^2 |T|^2 |T-1/2|^2 \, ,
\label{gen1m}
\eeq
by making the field redefinition $2T = e^{-2\chi/\sqrt{3}}$. The potential using (\ref{gen1m}) can be obtained from that using (\ref{gen1})
by making the SU(1,1) inversion transformation $T\to 1/(4T)$, see (\ref{invert}).

The second general form applies to either the generic fields $y_{1,2}$ or the `matter' field $\phi$.
The form of the potential is now 
\beq
\hat{V} \; = \; M^2 |\phi|^2 |1-\phi/\sqrt{3}|^2 \, ,
\label{gen2}
\eeq
or the equivalent for $y_{1,2}$. Incorporating the field-dependent factor in (\ref{Vhatphii}), we see that
this yields a potential that is independent of $\phi$ in the limit of large $\phi$, and hence also invariant asymptotically
under the dilatation transformation (\ref{alpha}). 
In this and similar cases, the appropriate field redefinition is 
\beq
(y_i,\phi) \; = \; \sqrt{3} \tanh \left( \frac{\chi}{\sqrt{3}} \right) \, .
\eeq
which yields (\ref{arctanh}) for the real part of $\chi$.
The Lagrangian now becomes
\begin{eqnarray}
{\cal L}  & = & {\rm sech}^2((\chi-\chi^*)/\sqrt{3}) \left[ | \partial_\mu \chi|^2  \right. - \\
& & \, \, \left. 3 M^2 \left|\sinh(\chi/\sqrt{3}) \left(  \cosh(\chi/\sqrt{3})- \sinh(\chi/\sqrt{3})\right) \right|^2 \right]  \, . \nonumber
\end{eqnarray}
This is identical to the Lagrangian in (\ref{leff}) (after some manipulation of the
exponential and hyperbolic functions) and 
writing $\chi$ in terms of its real and imaginary parts:
$\chi = (x + iy)/\sqrt{2}$ we obtain the same Lagrangian shown in (\ref{leff2})
For $\langle y \rangle = 0$, we again recover the potential (\ref{r2pot}) in terms of $x$.

We now exhibit some specific examples of SU(2,1)/SU(2) $\times$ U(1) no-scale inflationary models
within this general framework, noting correspondences to examples in the previous literature.

\subsubsection*{I. Example from~\cite{ENO6}}

This is based on the choice
\begin{equation}
W \; = \; M \left[ \frac{y_1^2}{2} \left(1+\frac{y_2}{\sqrt{3}} \right) - \frac{y_1^3}{3 \sqrt{3}} \right] \, ,
\label{W1}
\end{equation}
which is a Wess-Zumino (WZ) model for $y_1$ with an interaction term $y_1^2 y_2$.
In this case, even with
the assumption that $y_2$ is fixed so that $\langle y_2 \rangle = 0$,
$W$, $W_1$, and $W_2$ are all non-zero, and using (\ref{useful})
we obtain the effective potential
\begin{equation}
V \; = \; \frac{M^2 |y_1|^2 ~ |1 - y_1/\sqrt{3}|^2}{(1 - |y_1|^2/3)^2} \, ,
\label{V1}
\end{equation}
which is dilatation-invariant for large $y_1$ and precisely of the form (\ref{gen2}), and therefore yields
exactly the Starobinsky potential. Transforming back to the $(T, \phi)$ basis using
(\ref{Tphirewrite}), we obtain the following expressions for the K\"ahler potential and the
superpotential:
\begin{equation}
K \; = \; - 3 \ln \left(T + T^* - \frac{|\phi|^2}{3} \right) \, , \, W \; = \; M \left[\frac{\phi^2}{2} - \frac{\phi^3}{3 \sqrt{3}} \right] \, .
\label{ENO6}
\end{equation}
This is exactly the Starobinsky example of~\cite{ENO6}, in which the inflaton field is
identified as a `matter' field with the WZ superpotential, assuming that the modulus is fixed at $\langle T \rangle = 1/2$.

\subsubsection*{II. Reversed Example}

We now consider the reversed choice
\begin{equation}
W \; = \; M \left[ \frac{y_2^2}{2} \left(1+\frac{y_1}{\sqrt{3}} \right) - \frac{y_2^3}{3 \sqrt{3}} \right] ,
\label{W2}
\end{equation}
and assume that $y_1$ is fixed so that $\langle y_1 \rangle = 0$. 
Since this is exactly the same potential as Example I with $y_1$ and $y_2$ interchanged,
it again produces exactly the Starobinsky potential (\ref{r2pot}).
Performing
the transformation to the $T, \phi$ basis using
(\ref{Tphirewrite}) ({\em without interchanging $y_1$ and $y_2$}),  we obtain the same expression for the K\"ahler potential as in (\ref{ENO6}),
but the superpotential becomes
\begin{equation}
W \; = \; \frac{M}{4} (T - 1/2)^2 (1 + 10 T +2 \sqrt{3} \phi ) \, .
\label{Invert}
\end{equation}
This yields the effective potential
\begin{equation}
V \; = \; \frac{12 M^2  |T|^2 |T - 1/2|^2}{(T + T^*)^2}
\label{Vinvert}
\end{equation}
which is precisely of the form (\ref{gen1m})
and, making the transformation $T = e^{-\sqrt{2/3} x}/2$,
we see that this example also reproduces the Starobinsky potential,
but with the inflaton identified as the `modulus' field and with $\phi$ fixed at 0.

On the other hand, transforming $y_2 \to - y_2$ in (\ref{W2}),
we would obtain 
\begin{equation}
W \; = \; \frac{M}{4} (T - 1/2)^2 (5 + 2 T +2 \sqrt{3} \phi ) \, ,
\label{Invertm}
\end{equation}
which gives the asymptotically dilatation-invariant potential
\begin{equation}
V \; = \; \frac{3 M^2  |T - 1/2|^2}{(T + T^*)^2}
\label{Vinvertm}
\end{equation}
which is now precisely of the form (\ref{gen1})
requiring the transformation $T = e^{\sqrt{2/3} x}/2$.
Once $\phi$ is properly stabilized, these superpotentials both yield the 
same scalar potential for $Re~\chi$. 

\subsubsection*{III. Alternative Example~\cite{Cecotti,KLno-scale}}

Next we consider an example based on the superpotential
\begin{equation}
W \; = \; M y_1 y_2 (1 + y_2/\sqrt{3} ) \, ,
\label{Ex3}
\end{equation}
which yields
\begin{equation}
W_1 \; = \; M y_2 (1 + y_2/\sqrt{3} ) .
\label{Wone}
\end{equation}
If we assume that $\langle y_1 \rangle = 0$, so that $W, W_2 = 0$,
 ${\hat V}$ is particularly simple:
\begin{equation}
{\hat V} \; = \; |W_1 |^2 \; = \; M^2 |y_2|^2 | 1 + y_2/\sqrt{3} |^2 \, ,
\label{Vhat3}
\end{equation}
which is again of the form of (\ref{gen2}) (with $y_2 \to -y_2$)
and making the transformation $y_2 = - \sqrt{3} \tanh (x/\sqrt{6})$ reproduces the
Starobinsky potential again. Transforming to the $(T, \phi)$ field basis, we find that
\begin{equation}
W \; = \sqrt{3} M \phi (T - 1/2)
\label{W3}
\end{equation}
as in~\cite{Cecotti,KLno-scale}, and the potential is identical to that in the previous `reversed' case (\ref{Vinvert})
with the modulus $T$ associated with the inflaton.

As in the previous example, we could take $y_2 \to - y_2$ in (\ref{Ex3})
and find \begin{equation}
W \; = 2 \sqrt{3} M \phi  T (T - 1/2) \, ,
\label{W3invert}
\end{equation}
after the redefinition to the $(T, \phi)$ basis. 
Not surprisingly, this yields the same potential found in (\ref{Vinvert}).

\subsubsection*{IV. Alternative Reversed Example}

Consider the `reversed' version of the previous example (\ref{Ex3}), namely
\begin{equation}
W \; = \; M y_2 y_1 (1 + y_1/\sqrt{3} ) \; {\rm with} \; \langle y_2 \rangle = 0 \; ,
\label{Ex4}
\end{equation}
which is formally equivalent. However, when transformed to the $(T, \phi)$ field basis
it yields
\begin{equation}
W \; = \; M \left[\sqrt{3} (T^2 - 1/4) \phi + (T - 1/2) \phi^2 \right] \, .
\label{W4}
\end{equation}
In this case, with $\langle T \rangle = 1/2$, $W=W_\phi=0$ and $W_T = \sqrt{3} \phi -\phi^2$ 
and hence it yields the same potential as the first example (\ref{V1}) (with $\phi \to -\phi$).

These few examples demonstrate that no-scale Starobinsky models discovered previously~\cite{ENO6,Cecotti,KLno-scale}
are not unique. Indeed we have written down 4 explicit and different theories
which each lead to the Starobinsky model of inflation when either $\phi$ or $T$ (or $y_1$ or $y_2$) are properly stabilized. 
We do not attempt here a complete categorization of such models, but we do display some
classes of generalizations. 

\subsubsection*{Some Generalizations}

We consider first a generalization of example \{1\} above:
\begin{equation}
W \; = \; M \left[ \frac{y_1^2}{2} \left(1+\frac{y_2}{\sqrt{3}} \right) - \frac{y_1^3}{3 \sqrt{3}} \right]+ g(y_1, y_2) \, ,
\label{W1g}
\end{equation}
where the extra term $g(y_1, y_2)$ is chosen so that $g(y_1, 0), \partial g/\partial y_1 (y_1, 0)$
and $\partial g/\partial y_2 (y_1, 0) = 0$, one such example being 
\beq
\{1{\rm g}\} \; \; \; \; \; \; \; \; \; \; \; \; \; \; \; \; \; \; \;
g(y_1, y_2) = (y_2/\sqrt{3})^n: n > 1.
\eeq
It is clear that under these assumptions the potential will be identical to that in example \{1\} when $\langle y_2 \rangle = 0$.
If we consider the same model in the $(T, \phi)$ frame, the effective superpotential receives a contribution
\begin{equation}
\Delta W \; = \;  \left[ \frac{(T - 1/2)^n 2^{(n-3)}}{(2T + 1)^{(n-3)}}\right] \, ,
\label{DeltaW}
\end{equation}
which makes no contribution to the effective potential $V$ when one fixes $\langle T \rangle = 1/2$.
Alternatively, one could choose 
\begin{equation}
\{2{\rm g}\} \; \; \; \; \; \; \; \; \; \; \; \; \; \; \; \; \; \; \;
g(y_1, y_2) \; = \; \left( \frac{y_2}{\sqrt{3}} \right)^n y_1: \; \; n > 1 \, ,
\label{otherg}
\end{equation}
in which case
\begin{equation}
\Delta W \; = \; \left[ \frac{(T - 1/2)^n 2^{(n-2)} \phi}{(2T + 1)^{(n-2)}}\right] \, .
\label{DeltaW2}
\end{equation}
Making the choice $n = 2$ yields
\begin{equation}
\Delta W \; = \;  (T - 1/2)^2 \phi \, ,
\label{n=2}
\end{equation}
which is related to the previous examples of~\cite{ENO6} and~\cite{Cecotti}.

One final simple example starts with the superpotential (\ref{Ex3}) and adds the function
$g = M y_1^2 y_2/\sqrt{3}$, which is the simplest generalization of type 2g. 
In this case, in the $(T,\phi)$ basis we have
\beq
W =  \sqrt{3} M \phi  (1+ \phi /\sqrt{3}) (T-1/2) \, ,
\eeq
where $T$ is assumed fixed and $\phi$ is the inflaton.
This superpotential is of the form (\ref{W3}) with an additional factor $(1+\phi/\sqrt{3})$ but
still results in the Starobinsky potential. 
Clearly one can generate yet other examples by reversing $y_1$ and $y_2$ in all of the
generalization discussed above.

It is possible to generalize in similar ways the other specific examples give above,
but we do not go into details here. The key observation is that, within the
framework of SU(2,1)/SU(2) $\times$ U(1) no-scale supergravity and, {\it a fortiori}
models containing it, there are many ways to obtain an effective inflationary
potential identical with that in the Starobinsky model. In some of these cases,
the inflaton is identified with a modulus field $T$ as might appear in a
generic string compactification, in others it is identified with
a `matter' field $\phi$. There is no fundamental distinction between these at the
level of the coset structure and the K\"ahler potential. However, the ways these fields
appear in string compactifications are different, with very different forms of
superpotential, as seen already in the original analysis of~\cite{Witten} where
the superpotential for the matter fields was related to gauge interactions in
ten dimensions.

\section{Stabilizing the Modulus Field in SU(2,1)/SU(2) $\times$ U(1) No-Scale Supergravity}

Up until now, we have tacitly assumed that three of the four real components of 
the two complex fields $(y_1, y_2)$ or $(T, \phi)$ have been been stabilized.
Achieving this field stabilization is a generic issue in such models with two more more
complex scalar fields. If one component is interpreted as the inflaton field, with a value
that slides down the effective (Starobinsky) potential during the inflationary epoch,
how may the other fields be fixed, or at least constrained so as not to spoil the
inflationary dynamics? In the context of string compactifications, this
is manifested as the problem of stabilizing moduli fields. In this Section
we give examples of mechanisms capable of fixing the `modulus' $T$ or `matter' field $\phi$
in examples where the inflaton is identified with the `matter' field $\phi$  or `modulus' $T$ respectively.
The stabilization mechanisms we present here are by no means unique,
and are not necessarily motivated by deeper theoretical considerations, but they do serve
as existence proofs.

Let us first consider Example I  from Section 3.
In this case we assumed that $\langle y_2 \rangle =0$, so that the dynamics of the rolling 
inflaton ($y_1$) is given by the Starobinsky potential (\ref{r2pot}) as determined by
the K\"ahler potential (\ref{K21symm}) with superpotential (\ref{W1}),
but there are two possible problems.
1) The potential may not be stabilized in the two $y_2$ directions (real and imaginary) when $y_1$ (the inflaton)
is at its minimum.
2) While the $y_2$ direction is stabilized when $y_1 \ne 0$, i.e., during inflation, 
its real part ${y_2}_R$ has a non-zero expectation value, though ${y_2}_I = 0$. 
Although the shift in ${y_2}_R$ is relatively small, it might be enough to perturb the inflationary dynamics of $y_1$.

Both of these issues have relatively simple solutions.
Tackling first problem 2): the shift in ${y_2}_R$ can be made sufficiently small if a higher-order term is added to the 
K\"ahler potential, 
\beq
K \; = \; - 3 \ln \left(1 - \frac{|y_1|^2 + |y_2|^2}{3} +  \frac{|y_2|^4}{\Lambda^2} \right) \, ,
\label{K1g}
\eeq
where $\Lambda$ is a mass scale assumed to be smaller than the Planck scale:
$\Lambda \la 0.3 M_P$ is sufficient to restore the inflationary trajectory of $y_1$.
Concerning problem 1): a mass term can be generated for $y_2$ at $y_1 = 0$ by adding the simplest generalization \{1g\} above,
i.e., taking 
\begin{equation}
W \; = \; M \left[ \frac{y_1^2}{2} \left(1+\frac{y_2}{\sqrt{3}} \right) - \frac{y_1^3}{3 \sqrt{3}} +b \frac{y_2^2}{3} \right]\, .
\label{W1gstab}
\end{equation}
While this additional term leaves $V(y_1)$ unaffected for $y_2 = 0$, it provides mass
terms for both the real and imaginary scalar components of $y_2$ proportional to the coupling $b$.

One can rewrite this theory in terms of a modulus $T$ and inflaton $\phi$ as in (\ref{ENO6})
and derive the corresponding correction terms. Alternatively, one can start with (\ref{ENO6})
and stabilize the theory in terms of these fields.
In this case, one can take the example proposed first in~\cite{EKN3} for stabilizing moduli, 
and consider the K\"ahler potential~\footnote{The term $d(T-T^*)^4$, which was not included in \cite{EKN3},
is included here to stabilize the imaginary part of $T$, while the real part is stabilized by $(T+T^*-1)^4$.}
\beq
K \; = \; - 3 \ln \left(T + T^* - \frac{|\phi|^2}{3} + \frac{(T+T^*-1)^4 + d(T-T^*)^4}{\Lambda^2} \right) ,
\label{ekn}
\eeq
where $\Lambda$ is again a mass scale somewhat smaller than the Planck scale, and $d$ is a parameter
that breaks the invariance of the no-scale K\"ahler potential under the imaginary translations (\ref{beta}),
and allows the masses of the real and imaginary parts of $T$ to differ: we will set $d=1$.  To obtain
a non-zero mass for T, it is sufficient to add a constant to the superpotential, which generates the gravitino mass or as in (\ref{W1gstab}), we can add an explicit mass term of the form
$b (T-1/2)^2$ to the superpotential.

\begin{figure}[h!]
\vskip -1.5in
\includegraphics[scale=.5]{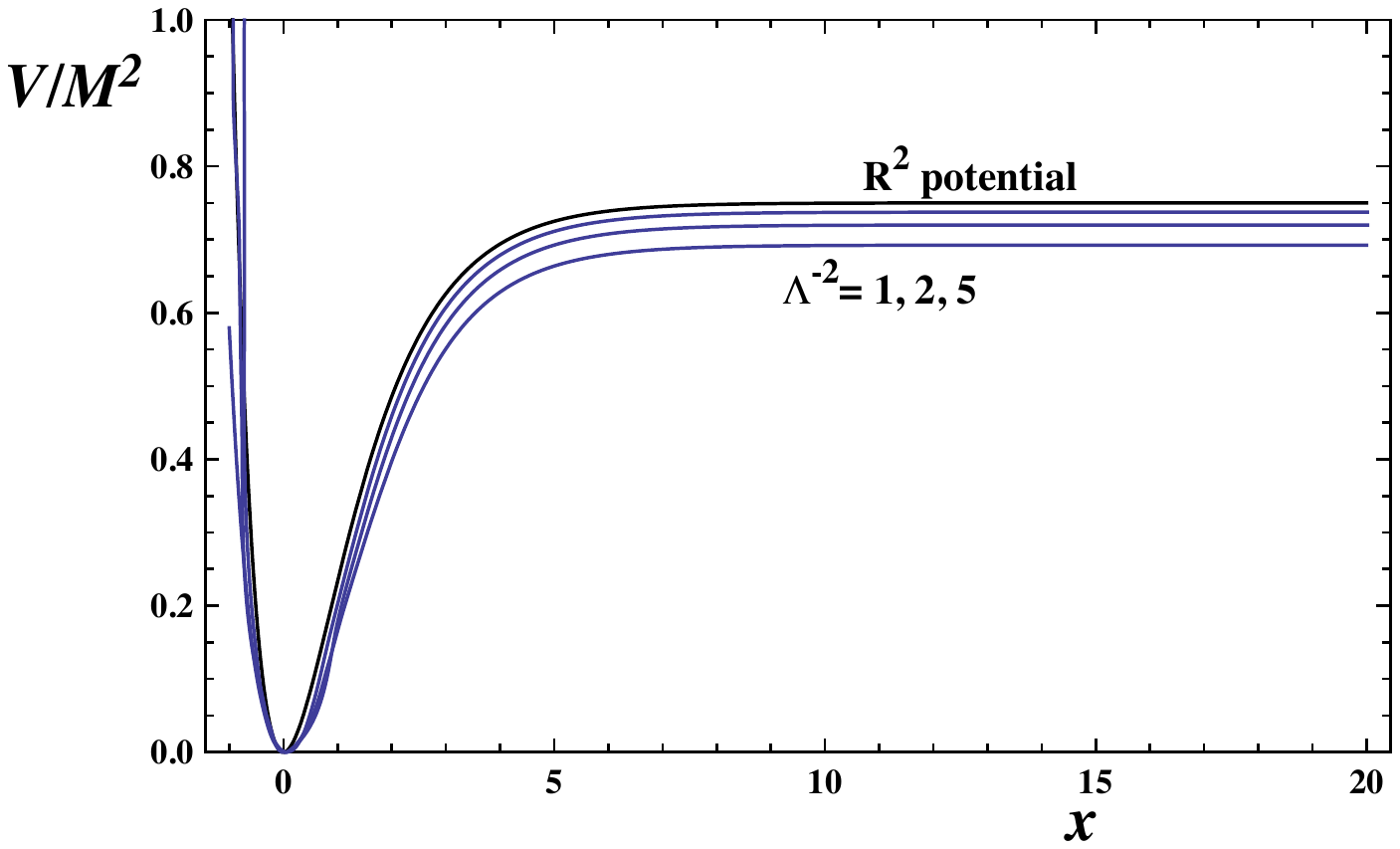}
\hskip -1.4in
\includegraphics[scale=.5]{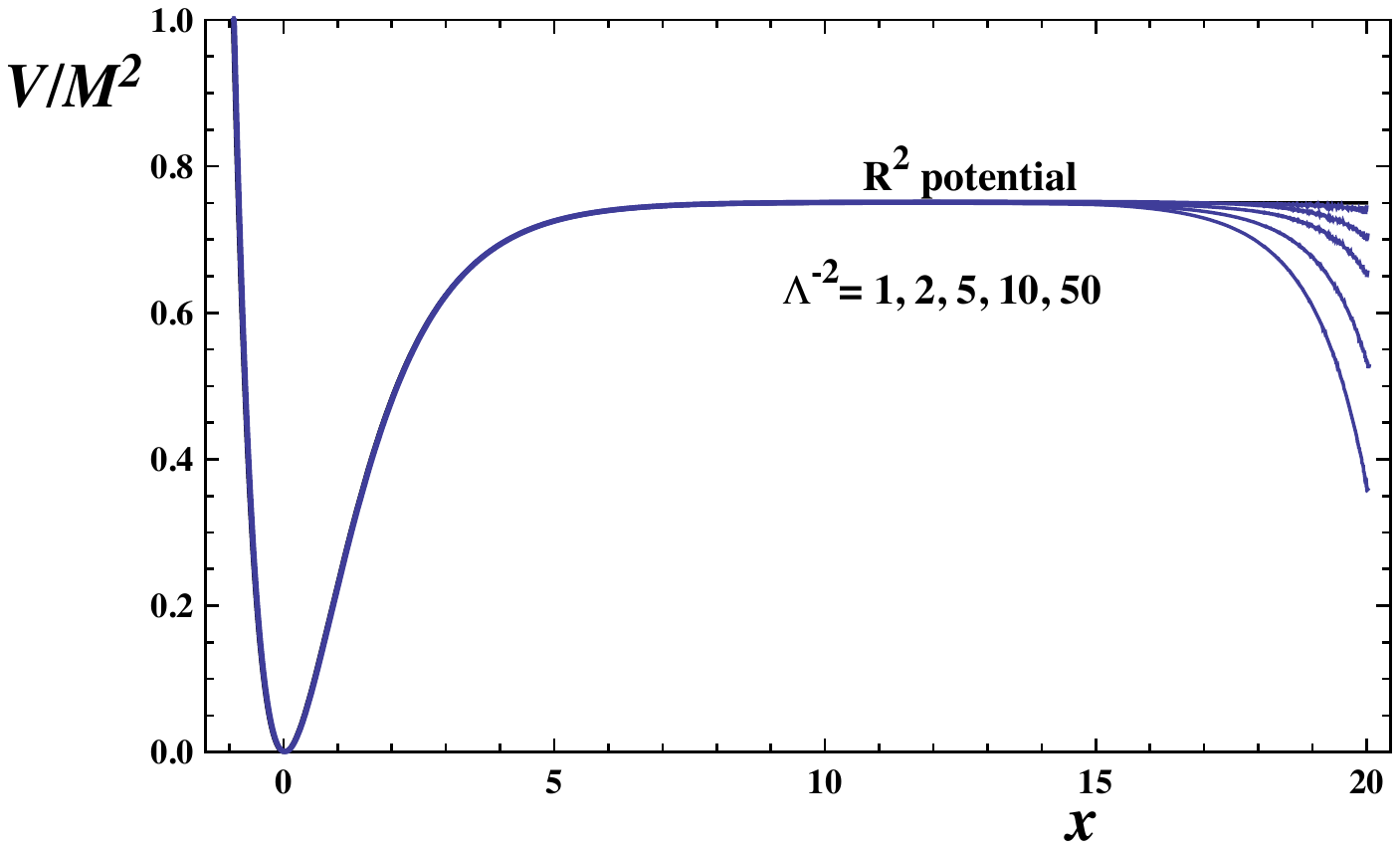}
\vskip -1.2in
\caption{\it The potential $V(x)$ evaluated at $\langle t  (x) \rangle$ for several  choices of  $\Lambda^2$ 
in Planck units, as indicated. For $\Lambda^2 \la 0.02$, the potential is indistinguishable from the potential in (\ref{r2pot}). In the left panel, a constant was added to the superpotential, while in the 
right panel, an explicit mass term was added.}
\label{staro}
\end{figure}

In the absence of the stabilizing term (\ref{ekn}), 
the potential in terms of the real parts of $\phi$ and $T$, takes the form
\beq
V = \frac{3 M^2 (1-\tanh(x/\sqrt{6}))^2 \tanh^2(x/\sqrt{6})}{(t-\tanh^2(x/\sqrt{6})^2} \, ,
\label{xt}
\eeq
where $x$ is the real part of the canonical field associated with $\phi$ (as in (\ref{arctanh}))
and  here, $Re~T = t/2$.
However, this potential gives no reason to suppose that $t$ will be fixed at 1,  the value
needed to recover the Starobinsky potential.
In the presence of the additional term in (\ref{ekn}), $\Lambda = {\cal O}(1)$ is sufficient to 
fix $t$ very close to 1, and produces a potential very similar to that of the Starobinsky model.
In Fig. \ref{staro} we display the resultant scalar potential for $x$. For each value of $x$, $t$ is evaluated at its local minimum near $t=1$. In the left panel we show the potential  for three choices of the mass scale
$\Lambda^{-2} = 1, 2$, and 5 with the constant in the superpotential chosen so
that $m_{3/2}=10^{-6}$. When $\Lambda^{-2} = 10$, it differs from the pure $R + R^2$ model potential by less than 1\% at $x = 20$,
and for  $\Lambda^{-2} = 50$ the difference is less than 0.2\%.
In the right panel, we show the potential when the mass term is added (instead of a constant),
with $b=10^{-6}$ for five choices of the $\Lambda^{-2} = 1, 2, 5, 10$ and 50. 

\begin{figure}[h!]
\centering
\includegraphics[scale=1]{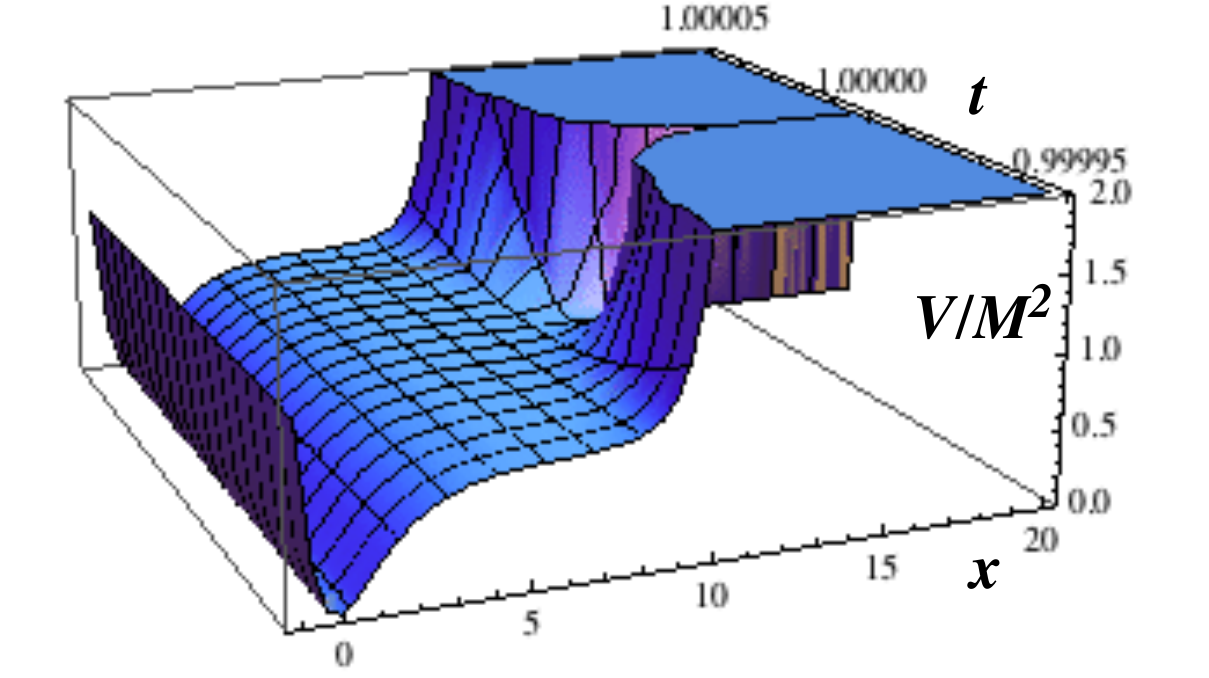}
\caption{\it The scalar potential in the $(Re~T, Re~\phi)$ space using the coordinates defined in connection with
(\protect{\ref{xt}}), for $Im~T  = Im~\phi  = 0$.}
\label{3D}
\end{figure}

A three-dimensional view of the potential in the $(Re~T, Re~\phi)$ space is shown in Fig.~\ref{3D} for
$\Lambda^{-2} = 50$, where we see the strong stabilization at large $x$.  
Although the potential appears to flatten in the $Re~T$ direction,
it remains stabilized at all values of $x$. The constant in the superpotential was chosen so that $m_{3/2}=10^{-6}$,
and the curvature in the $t$ direction is imperceptible when $x \to 0$
on the scale of the figure. We also note that $m_t \propto \mathcal{O}(10) m_{3/2}/\Lambda$, and hence is
hierarchically larger than the gravitino mass. When a mass term is added to the superpotential instead of a constant,
the choice $b=10^{-6}$ would yield an almost identical potential.
The inflationary trajectory begins at moderate or large $x$ and emerges from the crack in the potential
on the right side of the figure. 

Clearly the theory described by the superpotential (\ref{W2}) would be stabilized in an identical manner as 
described in (\ref{K1g}) and (\ref{W1gstab}).  However, in this case when $y_1$ and $y_2$ are reversed, 
it is the $T$ field that plays the role of the inflaton when fields are transformed to the $(T, \phi)$ basis.
We do not discuss stabilization for this case, but instead consider Example III from the previous Section. 
In this case it is sufficient to add the stabilizing term to the K\"ahler potential alone, and we choose
\beq
K \; = \; - 3 \ln \left(1 - \frac{|y_1|^2 + |y_2|^2}{3} +  \frac{|y_1|^4}{\Lambda^2} \right) \, ,
\eeq
along with the superpotential given by (\ref{Ex3}).
The mass of $y_1$ is non-zero and proportional to $M$ when the inflaton ($y_2$ in this case) is at its minimum.
Thus no correction to $W$ is necessary.

It is interesting to note that, in this case, the point $y_1=0$ is always an extremum. However,
in the absence of the stabilizing term in $K$, it is a local maximum and therefore represents 
an instability, which is critical in this case. Turning on the $\Lambda$-dependent stabilization term increases
the curvature at $y_1 = 0$.  For $\Lambda^{-2} < 50$, the curvature is positive for all values of $x \la 10$
(where $x$ is the canonical field associated with $y_2$).
To extend to larger values of $y_2$, a smaller value of $\Lambda$ should be chosen.
We recall that inflation requires only that $x\ga 5$. Thus stabilization in this theory is relatively easy to achieve.
Writing this theory in the $(T, \phi)$ basis gives us the superpotential shown in (\ref{W3}),
and 
\beq
K \; = \; - 3 \ln \left(T + T^* - \frac{|\phi|^2}{3} +  \frac{8 |\phi|^4}{\Lambda^2 |1+2T|^2}  \right) \, .
\eeq
Had we started in the $(T, \phi)$ basis, we could have used a simpler form for the 
K\"ahler potential~\cite{KLno-scale}, namely
\beq
K \; = \; - 3 \ln \left(T + T^* - \frac{|\phi|^2}{3} +  \frac{ |\phi|^4}{\Lambda^2 }  \right) \, ,
\eeq
and obtained qualitatively similar results.  

Other examples discussed in the previous Section
can be stabilized with similar corrections, i.e., adding a $|\phi|^4$ term to $K$
for stabilizing fields like $(y_i, \phi)$ or by adding a $(T+T^*)^4$ term to $K$ for
stabilizing $T$ fields. 

\section{Exploring the Parameter Space of Starobinsky-Like Models}

We now consider some theoretical possibilities for constructing
within the no-scale framework models 
that resemble the original Starobinsky model but make 
predictions for the CMB observables that can in principle be
distinguished experimentally, while lying within the range allowed by
present observations. 

We recall the Starobinsky potential can be expressed in
the simple form
\begin{equation}
V \; = \; A \left( 1 - e^{-Bx} \right)^2 \, ,
\label{StaroAB}
\end{equation}
where $x$ is a canonically-normalized field, the value of $A$ fixes the magnitude of the scalar density
perturbations, and $B = \sqrt{2/3}$. We note that the potential (\ref{StaroAB}) is positive
semi-definite, vanishing iff $x = 0$, but observe that the inflationary predictions are
derived in the large-field regime where the constant and leading term in $e^{-Bx}$
are dominant. The behaviour of the potential away from this large-field regime is
irrelevant for the inflationary predictions we discuss here.

In~\cite{ENO6} we considered a no-scale model in the $(T, \phi)$ frame with a Wess-Zumino superpotential (\ref{WZW}).
In terms of the canonically-normalized real component of the field $x: Re\phi \equiv \sqrt{3c} \tanh (\sqrt{2/3}x)$
where we define $c \equiv 2 \langle Re T \rangle$ and $\mu \equiv {\hat \mu}/\sqrt{c/3}$,
we found the effective potential
\begin{equation}
V = \mu^2\left |\sinh(\sqrt{2/3}x) \left( \cosh(\sqrt{2/3}x)-\frac{3\lambda}{\mu} \sinh(\sqrt{2/3}x) \right) \right|^2 \, .
\label{WZV}
\end{equation}
It is clear that when one makes the particular choice $\lambda = \mu/3$,
the potential (\ref{WZV}) is of the form (\ref{StaroAB}). However, when
$\lambda \ne \mu/3$ the potential (\ref{WZV}) grows exponentially
for large $|\chi|$, as seen in Fig.~1 of~\cite{ENO6}. In the region of interest where $\lambda \sim \mu/3$,
the values of $V$ and $V^\prime$ do not differ much from the Starobinsky case (\ref{StaroAB}),
so the value of $\epsilon$ and hence $r$ are similar to those in the Starobinsky model,
increasing slightly as $\lambda/\mu$ decreases, as seen
Fig.~2 of~\cite{ENO6}. On the other hand, when
$\lambda < \mu/3$ there is an inflection point: $V^{\prime \prime}$ = 0 near the starting-point
of inflation, so that $\eta$ may very small and $n_s \sim 1$, as also seen in
Fig.~2 of~\cite{ENO6}.

Here we consider phenomenological generalizations of (\ref{StaroAB}) in which
\begin{equation}
V \; = \; A \left( 1 - \delta e^{-Bx} + {\cal O}(e^{-2Bx}) \right) \, ,
\label{StaroAlambdaB}
\end{equation}
with $\delta$ and $B$ treated as free parameters that  may deviate from the
Starobinsky values $\delta = 2$ and $B = \sqrt{2/3}$. In such a case, at leading order in the small quantity $e^{-Bx}$ one finds
\begin{eqnarray}
n_s & = & 1 - 2 B^2 \delta e^{-Bx} \, , \nonumber \\
r & = & 8 B^2 \delta^2 e^{-2Bx} \, , \nonumber \\
N_* & = & \frac{1}{B^2 \delta} e^{+ Bx} \, .
\label{predictions}
\end{eqnarray}
yielding the relations
\begin{equation}
n_s \; = \; 1 - \frac{2}{N_*} \; , r \; = \; \frac{8}{B^2 N_*^2} \, .
\label{relations}
\end{equation}
Requiring $N_* = 54 \pm 6$ yields the characteristic predictions $n_s = 0.964 \pm 0.004$, 
and the Starobinsky choice $B = \sqrt{2/3}$ yields $ r = 12/N_*^2 = 0.0041^{+0.0011}_{-0.0008}$.
These predictions are explicitly independent of $\delta$.

The question then arises how one could deviate from the characteristic Starobinsky
prediction for $r$, which would require a different value of $B$. One possibility is
to consider models with multiple moduli that share the no-scale property 
$({\partial K}/{\partial  \phi^i}) K^i_{j^*} ({\partial K}/{\partial  \phi^*_j}) = 3$:
\begin{equation}
K \; \ni \; - \Sigma_i \, N_i \, \ln (T_i + T_i^*) : \; \; N_i > 0, \; \; \Sigma_i \, N_i = 3 \, . 
\label{multimoduli}
\end{equation}
Such models have similar properties under the SU(2,1) transformations (\ref{beta},\ref{alpha},\ref{tau},\ref{invert})
as the original no-scale model (\ref{K3}).
If one identifies the inflaton with the the modulus field $T_i$ whose logarithmic coefficient is $N_i$,
the corresponding transformation to a canonically-normalized field is $T_i \sim e^{\sqrt{2/N_i}x}/2$.
We have not made a detailed study of models based on this identification, but it is easy to find
modifications of the $N_i = 3$ superpotential (\ref{WZW}) that yield an inflaton potential of
the form (\ref{StaroAB}) but with
\begin{equation}
B \; = \; \sqrt{\left(\frac{2}{N_i}\right)} \, ,
\label{generalB}
\end{equation}
so that
\begin{equation}
r \; = \; 
\frac{4 N_i}{N_*^2} \, .
\label{generalr}
\end{equation}
The sample models we have found are not very attractive, but they do make the point
that no-scale supergravity could accommodate a Starobinsky-like model with a significantly
different value of $r$. We defer the detailed exploration of such possibilities for possible
future work.

Realistically, the leading alternative to the single-modulus case with $N_i = 3$
may be a three-modulus case with $N_i = 1$, in which case $r$ would be a factor of 3 smaller than in the
Starobinsky model. Within the class of no-scale models discussed here,
a measurement of $r$ might eventually provide some observational information
on the form of string compactification.

\section{Conclusions}

We have shown in this paper that the connection between the Starobinsky model of inflation and no-scale
supergravity found in~\cite{ENO6} is both deeper and broader than the example given there.
As discussed in Section~2 of this paper, the connection is deeper in the 
sense that the form of the kinetic energy for the scalar field in the
conformal reformulation of $R + R^2$ gravity after Weyl rescaling (\ref{Cecotti4})~\cite{whitt} is identical~\cite{Cecotti}
to that for the real part of the `modulus' field in no-scale supergravity (\ref{no-scaleLKE})~\cite{no-scale},
which is a basic feature of its K\"ahler geometry,
reflecting the common dilatation invariance (\ref{alpha}) of these kinetic terms. Because of this
underlying geometric origin of the connection, it is also broader as discussed in Section~3,
in the sense that there is considerable freedom of choice in the form of superpotential that
reproduces the Starobinsky inflationary potential (\ref{r2pot}).

The no-scale framework is, however, more general than the specific Starobinsky model, opening up
the possibility of studying a more general class of models within which Starobinsky is embedded.
This in turn provides a phenomenological context where one can explore the extent to which
observational data push cosmological models into Starobinsky's arms. Concretely, no-scale
models offer many ways to generalize the Starobinsky model by varying the choice of
superpotential, and a further discrete set of choices for the K\"ahler potential. A one-parameter
set of options for varying the superpotential was explored in~\cite{ENO6}, namely varying the
ratio of the two parameters $\hat \mu$ and $\lambda$ in the superpotential of the Wess-Zumino
model (\ref{WZW}). As was pointed out in~\cite{ENO6}, whereas the particular choice $\lambda = \mu/3$
(where $\mu = {\hat \mu}/\sqrt{c/3}: c = \langle T + T^* \rangle/2$) reproduces the Starobinsky model,
whereas models with $\lambda \ne \mu/3$ generalize it. As was discussed in~\cite{ENO6}, the
range of $\lambda/\mu$ that leads to inflationary models compatible with experiment is very limited,
essentially by the observational limit on $n_s$. For $N_* = 55$, only the range
\begin{equation}
0.33332\;  < \; \lambda/\mu \; < \; 0.33335
\label{range}
\end{equation}
is compatible with the Planck data at the 68\% CL, increasing to the range $(0.33331, 0.33337)$
at the 95\% CL. The Planck constraint on $n_s$ is likely to be the most important constraint on a
wide range of no-scale models with modified superpotentials. 

As was pointed out in the previous
Section, on the other hand, modifying the coefficient of the logarithm in the no-scale
K\"ahler metric would, in general, reduce substantially the Starobinsky prediction for $r$.
The latter lies well below the current observational sensitivity, though there are proposals for
projects with the sensitivity to establish a signal at the level of the Starobinsky prediction~\cite{Prism}.
A measurement at this level would not distinguish between $R + R^2$ gravity and the simplest
no-scale possibilities. However, a measurement below this level could provide non-trivial
information about the no-scale K\"ahler potential and how the inflaton field is embedded in it,
opening a new frontier in no-scale phenomenology. Conversely, a measurement of $r$
substantially larger than the $R + R^2$ prediction would be a strike against this no-scale framework.

The Planck data raise significantly the stakes in inflationary cosmology, with many simple models
now being disfavoured at the 68 or 95\% CL, e.g., $\phi^n: n \ge 2$ monomial models, while
the $R + R^2$ model remains viable. When
exploring the extended parameter space of more complicated models, it is desirable to follow
some guiding principles motivated by other physical considerations. One example is supersymmetry,
presumably in its local form, i.e., supergravity. Within this general framework, we consider
no-scale supergravity models to be the best motivated, since they open up the
possibility of determining dynamically a hierarchy of mass scales and emerge naturally in
compactifications of string theory. It is remarkable that no-scale models accommodate naturally
the $R + R^2$ model, while offering generalizations that can be probed by future CMB
experiments.

\section*{Acknowledgements}

J.E. thanks Nick Mavromatos
for discussions, and K.A.O. thanks Renata Kallosh, Nemanja Kaloper, Andrei Linde and
Misha Voloshin for discussions. The work of J.E. was
supported in part by the London Centre for Terauniverse
Studies (LCTS), using funding from the European
Research Council via the Advanced Investigator Grant
267352. The work of D.V.N. was supported in part by the
DOE grant DE-FG03-95-ER-40917. The work of K.A.O.
was supported in part by DOE grant DE-FG02-94-ER-40823 at the University of Minnesota.


\end{document}